\documentclass{emulateapj}

\usepackage{natbib}

\newcommand{\be}{\begin{equation}}
\newcommand{\ee}{\end{equation}}\newcommand{\beq}{\begin{eqnarray}}
\newcommand{\eeq}{\end{eqnarray}}

\newcommand{\halpha}{\mbox{H\hspace{0.2ex}$\alpha$}}

\slugcomment{ApJ Letters in press}

\shortauthors{Hansteen, et al. }
\shorttitle{Dynamic Fibrils and Magnetoacoustic Shocks}

\begin{document}
\title{Dynamic Fibrils Are Driven by Magnetoacoustic Shocks} 

\author{V. H. Hansteen\altaffilmark{1}}
\affil{Institute of Theoretical Astrophysics, University of Oslo, PO
  Box 1029 Blindern, 0315 Oslo, Norway}
\altaffiltext{1}{Also at: Center of Mathematics for Applications, 
University of Oslo, P.O.~Box~1053, Blindern, N--0316 Oslo, Norway}
\email{viggo.hansteen@astro.uio.no}

\author{B. De Pontieu}
\affil{Lockheed Martin Solar and Astrophysics Lab, 3251 Hanover St.,
  Org. ADBS, Bldg. 252, Palo Alto, CA 94304, USA}
\email{bdp@lmsal.com}

\author{L. Rouppe van der Voort\altaffilmark{1}}
\affil{Institute of Theoretical Astrophysics, University of Oslo, PO
  Box 1029 Blindern , 0315 Oslo, Norway}

\author{M. van Noort}
\affil{Institute for Solar Physics of the Royal 
Swedish Academy of Sciences, AlbaNova University Center, 106 91 Stockholm, Sweden}

\author{M. Carlsson\altaffilmark{1}}
\affil{Institute of Theoretical Astrophysics, University of Oslo , PO
  Box 1029 Blindern, 0315 Oslo, Norway}

\begin{abstract}
  The formation of jets such as dynamic fibrils, mottles, and spicules
  in the solar chromosphere is one of the most important, but also
  most poorly understood, phenomena of the Sun's magnetized outer
  atmosphere. We use extremely high-resolution observations from the
  Swedish 1-m Solar Telescope combined with advanced numerical
  modeling to show that in active regions these jets are a natural
  consequence of upwardly propagating slow mode magneto\-acoustic
  shocks. These shocks form when waves generated by convective flows
  and global p-mode oscillations in the lower lying photosphere leak
  upward into the magnetized chromosphere. We find excellent agreement
  between observed and simulated jet velocities, decelerations,
  lifetimes and lengths. Our findings suggest that
  previous observations of quiet sun spicules and mottles may also be
  interpreted in light of a shock driven mechanism.
\end{abstract}

\section{Introduction}

The solar chromosphere is sandwiched between the surface, or
photosphere, and the hot and tenuous outer corona. This highly
structured region, on average 2000~km thick, is constantly perturbed
by short lived (3 -- 10 minutes), jet-like extrusions that reach
heights of 2000 -- 10000~km above the photosphere. These thin jets are
formed in the vicinity of photospheric magnetic field concentrations.
Until recently, their small size and short lifetimes have made
detailed analysis difficult \citep[]{Beckers1968,Suematsu+etal1995},
which has led to a multitude of poorly constrained theories of their
formation \citep[]{Sterling2000}. In addition, there has been
considerable confusion about the relationship between spicules at the
quiet Sun limb, mottles observed on the quiet Sun disk, and dynamic
fibrils (DFs) found in the vicinity of active region plage
\citep[]{Grossmann-Doerth+Schmidt1992}, although the similarity in
many of their properties strongly suggests some of these phenomena are
related \citep{Tsiropoula+etal1994}. We focus on
observations of DFs (\S 2), compare them to advanced numerical
simulations (\S 3), report on regional differences of DF
properties (\S 4), and finish with a comparison to quiet Sun jets (\S 5).

\section{Observations of Dynamic Fibrils}

The recent advent of the Swedish 1-m Solar Telescope
\citep[SST,][]{scharmer2003SST} and advances in post-processing
techniques \citep[]{vanNoort05MOMFBD} have allowed us to obtain an
unprecedented, diffraction-limited (120~km) 78 minute long time series
of the chromosphere as imaged in the core of H$\alpha$ (656.3~nm) at a
cadence of 1~s \citep[]{vanNoort+Rouppe2006}.  
These data, taken on 4 October 2005, resolve for the
first time the spatial and temporal evolution of DFs, in particular the properties 
of the 257 DFs chosen for this study. While DFs have varying
lifetimes, lengths and widths, a typical fibril rises rapidly to a
maximum length in 1.5--3 minutes and recedes in a similar time along
the same, relatively straight path, presumably parallel to the
direction of the magnetic field (Fig.~1, top panels). Typical fibril 
lifetimes are between 120 and 650 s, with an average of
290 s. DFs display some internal structure (e.g.  at $t=139$ s
and $t=186$ s in top panels of Fig.~1): many DFs do not rise and fall
as a rigid body but rather show phase and amplitude variations in
velocity at various positions away from the fibril axis.  Despite the
substructure, fibrils are thin, with widths ranging from the
diffraction limit of 120~km to 700~km. The maximum projected extent is
usually relatively modest, ranging from 400 km to 5,000 km with an
average of 1,250 km. Both the lifetimes and lengths of DFs are in the
lower range of values reported for quiet Sun spicules or mottles
\citep{Beckers1968}.

It has been a subject of significant debate whether quiet Sun jets such
as spicules or mottles follow paths that are parabolic, ballistic (i.e.,
solar gravity), or constant velocity
\citep[e.g.,][]{Beckers1968,Nishikawa1988,Suematsu+etal1995}. The high
cadence and high spatial resolution of our observations reveals that 
most active region DFs observed in
\halpha\ line core follow almost perfect parabolic paths. We have therefore 
fit the paths to all 257 DFs outlined in \halpha\ linecenter with parabolas, 
and use these fits to determine velocities and decelerations. DF motions 
may be summarized by an initial
impulsive acceleration, after which the top of DFs are subjected to a
constant deceleration throughout their lifetime (Fig.~2, top panel). At
the beginning of the ascending phase, the velocity of the fibril top is
supersonic, with an average (line-of-sight projected) value of 18~km/s
and a range from 10 to 35~km/s for the 257 DFs. The velocity of the
fibril top then decreases linearly with time until it reaches a maximum
downward velocity that is roughly equal in amplitude to the initial
upward velocity.  The average projected deceleration is 73~m/s$^2$ with
a significant spread, ranging from 20 to 160~m/s$^2$. These projected
values are significantly smaller than the (downward) solar gravitational
acceleration of 274~m/s$^2$. In principle, the deceleration suffered by
DFs along their path could be much larger than the projected
deceleration, if their path is close to the line  of sight. 
We find that the projection angles
necessary to obtain solar gravity (or its component along the field) are
not compatible with the visual appearance
of the large scale topology of the active region (Fig. 4, top panels).

\begin{figure}
\epsscale{1.0}
\plotone{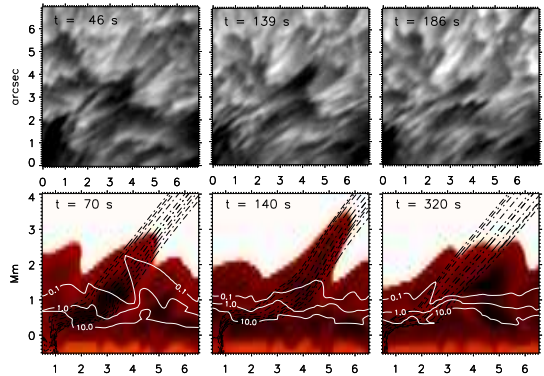}
\caption{Temporal evolution of dynamic fibrils from H$\alpha$
  linecenter observations at the Swedish 1-m Solar Telescope (top
  panels), and from numerical simulations (bottom panels). The
  observations show a dark elongated feature with an upper
  chromospheric temperature of less than 10,000 K rise and fall within 4
  minutes. The bottom panels show the logarithm
  of the plasma temperature T, set to saturate at log T = 4.5,
  from numerical simulations covering the upper convection zone
  (z$<$0) up through the corona (white region at the top). 
  The horizontal scale is arcsec in the top panels, and Mm in the bottom panels.
  The vertical scale has its origin at the photosphere (optical depth $\tau_{500}=1$). 
  Contours of plasma $\beta$ are drawn in
  white where $\beta=0.1, 1$ (thicker), $10$. The plasma $\beta$ is
  the ratio of the gas pressure to the
  pressure exerted by the magnetic field. 
  }
\label{fig1}
\end{figure}

\section{Numerical Simulations of Dynamic Fibrils}

We find features with similar properties in recent advanced
two-dimensional numerical simulations (Fig.~1, bottom panels). These
simulations model the evolution of a radiative MHD plasma, and, for
the first time, span the entire solar atmosphere from the upper
convection zone to the lower corona, including the photosphere,
chromosphere and transition region
\citep[]{Hansteen2005}.  This model includes non-grey, non-local thermodynamic
equilibrium radiative transport in the photosphere and chromosphere
\citep[]{Nordlund1982,Skartlien2000}, and optically thin radiative
losses as well as magnetic field-aligned heat conduction in the
transition region and corona. We used an average field strength of 
about 100 G, with the field clumping up to 1 kG in downdraft regions
in the photosphere. In these models, acoustic waves 
generated in or near the photosphere are found to propagate upward along the
magnetic field lines, form shocks in the middle chromosphere,
and lift the upper chromosphere several thousand kilometers along 
a front with a width of up to 1000~km.  The elevated plasma 
protrudes into the hot
corona (Fig.~1, bottom panels) and resembles the observed
DFs (Fig.~1, top panels). The duration of these simulated jets is of
order 3 - 5 minutes, with maximum lengths ranging from $500$~km to
several thousand km.

\begin{figure}
\epsscale{1.0}
\plotone{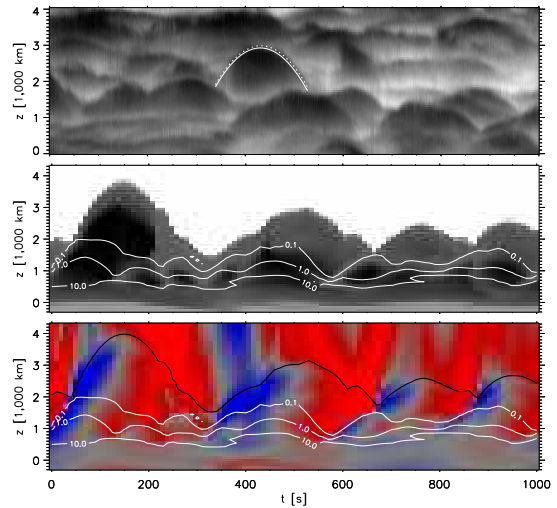} 
\caption{Space-time
    plots of: the height of several DFs from H$\alpha$ linecenter
    observations (top panel), the logarithm of the plasma temperature
    in numerical simulations (middle panel), and simultaneous plasma
    velocity from the same simulations (bottom panel).  Most of the
    DFs in the top panel follow a near perfect parabolic path, as
    illustrated for one DF by the two separate fits (full and dashed
    lines) used to derive parabolic parameters.  Parabolic paths with
    similar parameters are traced by fibril-like features in the
    simulations (middle panel). The vertical scale, plasma $\beta$
    contours and temperature range are the same as in the bottom
    panels of Fig.~1. Upward plasma velocities (blue in the bottom
    panel) show the upward propagation of the shocks that drive the
    fibril-like features. }
\label{fig2}
\end{figure}

Our numerical simulations explain why the DFs in our SST data have
parabolic paths with decelerations that are only a fraction of solar
gravity. In the simulations, fibril-like features are seen to follow
parabolic paths with properties very similar to those of the observed
DFs (Fig.~2, middle panel). These paths occur as a natural result of
shock wave driving. When a shock impacts the top of the chromosphere, the plasma
is catapulted upward at a velocity that exceeds the local sound speed
of $\sim$10 km/s. This agrees well with the lower cutoff at 10 km/s of
maximum velocities in the observed DFs. Shock waves generally have
velocity profiles in the form of 'N'- or 'sawtooth' shapes
\citep{Mihalas+Mihalas1984}. As a result, a plasma parcel passing
through a shock wave will first experience a sudden impulse in
velocity, followed by a gradual, linear deceleration as the shock
recedes --- this is consistent with a parabolic path.

The deceleration in these shock waves depends on the component of
solar gravity that is parallel to the magnetic field lines along which
the shock waves propagate. However, it also critically depends on the
period and amplitude of the shock waves. Even for vertical
propagation, our simulations show that the deceleration is less than
solar gravity. 'N'-shaped shock waves propagating at any angle to the
vertical, will, for a given period, show a steeper decline with time
(i.e., deceleration) in velocity for a greater shock strength.
Similarly, for a given amplitude, shock waves with shorter periods
show larger deceleration.  These dependencies explain why the observed
DFs have decelerations less than solar gravity. In addition, they
offer a natural explanation for an intriguing linear relationship
between the deceleration and maximum velocity of the DFs in our SST
data.  We find that DFs with a larger deceleration show a larger
maximum velocity (Fig.~3, top panel).  This previously unknown
correlation is well reproduced by the jets in our
simulations (Fig.~3, lower panel), for which we find a linear
relationship with a similar slope and similar range in values as for
the observed DFs.  The observed relationship and the close
similarities to simulations strongly suggest that DFs are driven by
chromospheric shock waves with strengths between Mach 1 and Mach
4.

\begin{figure}
\epsscale{1.0}
\plotone{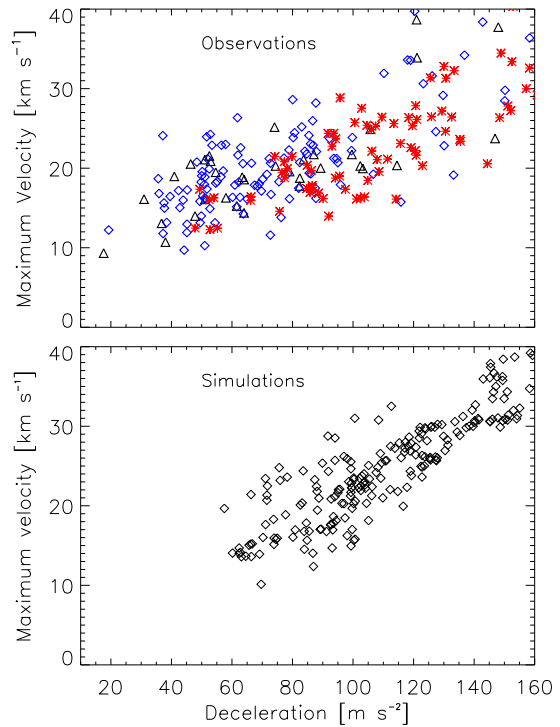} 
\caption{Top panel
    shows for the 257 DFs observed at the SST the maximum velocity
    versus deceleration, both corrected for line-of-sight projection,
    assuming that the fibril is aligned with the local magnetic field
    as deduced from potential field calculations. These corrections are
    uncertain, but as the same correction is applied to both quantities
    the correlation between them should not be affected by any errors 
    this correction introduces. DFs in region 1 and
    2 are indicated by, respectively, red stars and blue diamonds (see
    Fig.~4 for definition of regions). The same scatterplot (lower
    panel), based on analysis of fibril-like features in the numerical
    simulations, reveals that the simulations reproduce the observed
    correlation between these parameters, as well as reproducing the
    range in deceleration and maximum velocity.}
\label{fig4}
\end{figure}

\section{Regional Differences in Fibril Properties}

This scenario is further strengthened by the clear regional
differences of dynamic fibril properties. The field of view of the SST
observations contains a dense plage region (region 1, red circle in
Fig.~4), where the magnetic field is more vertically oriented, and a
less dense plage region (region 2, blue circle in Fig.~4), where the
field is more inclined away from the vertical. The dense plage region
has a prevalence of DFs with larger decelerations (typically
100~m/s$^2$, lower left panel of Fig.~4) than the region of less dense
plage where DFs have smaller decelerations (typically 50~m/s$^2$). In
addition, DFs are shorter, typically 1000~km, and have shorter
durations (3 minutes) in the dense plage region than in region 2,
where DFs are on average 2000~km and typically last for 5 minutes.
Since the regional differences imply high decelerations for short-lived
and short fibrils, projection effects cannot explain these differences.
If we assume that the low decelerations in region 1 are caused by a
different viewing angle than in region 2, this projection effect would
make the already longer fibrils in region 1 even longer than those in
region 2, which would make the regional differences more pronounced.
Moreover, differences in duration are independent of projection effects,
as are various correlations between deceleration, maximum length and
velocity.

\begin{figure} 
\epsscale{1.0}
\plotone{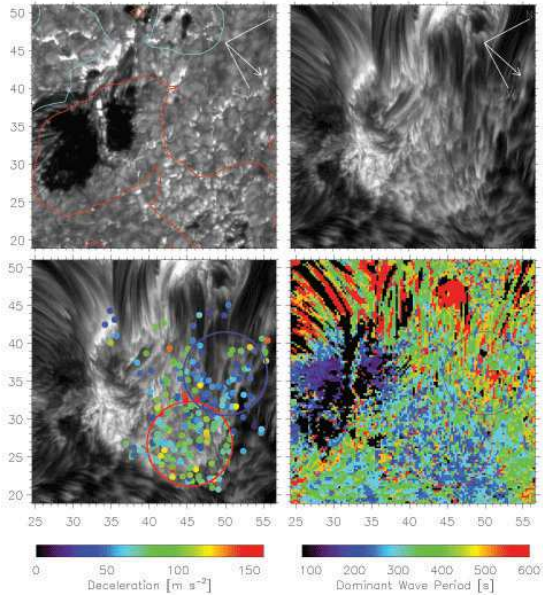} 
\caption{SST images
    taken in H$\alpha$ wideband (0.8~nm FWHM, upper left) and
    H$\alpha$ linecenter (upper right) of part of NOAA AR 10813, at
    heliocentric coordinates S7, E37. Top panels illustrate solar
    north and west, and the line-of-sight vector ($\theta=39\arcdeg$).
    Blue and red contours in the upper left panel outline positive and
    negative magnetic flux ($\pm$ 110 Mx/cm$^2$ in full-disk MDI
    magnetogram). Minor tickmarks in arcseconds. The DFs are
    predominantly observed in the unipolar plage region to the lower
    right of the sunspot.  The longer, more static and heavily
    inclined fibrils, visible at, e.g., x=30'' and y=45'' are not the
    subject of this paper. The lower left panel shows an H$\alpha$
    image superposed with dots, color-coded to illustrate fibril
    deceleration, at the location of each of the 257 DFs. The lower
    right panel shows the wave period containing the highest number of
    wavepackets with significant power in H$\alpha$ linecenter.
    Circles in the botttom panels show regions containing strong plage
    (region 1, red) and plage with inclined field (region 2, blue).}
\label{fig3}
\end{figure}

We believe instead that the regional differences occur because the
solar atmosphere produces chromospheric shocks of varying strengths
and varying periods. The strengths and periods depend critically on
which disturbances propagate from the photosphere into the upper
atmosphere. The chromosphere acts to filter out upward-propagating
disturbances with periods that are longer than the local acoustic
cutoff period. In general, the acoustic cutoff period depends on the
inclination of the magnetic field lines to the vertical
\citep[]{Suematsu1990,DePontieu+etal2004}. Under conditions of
vertical magnetic field such as in the dense plage region (region 1 in
Fig.~4), this implies that the chromosphere is dominated by
oscillations and waves with periods at the acoustic cutoff of 3
minutes. Since these waves shock and drive the DFs, this explains why
DF lifetimes are about 3 minutes in region 1. Wavelet analysis of the
H$\alpha$ linecenter time series confirms that the upper chromosphere
is dominated by wave trains with periods of order 3 minutes (Fig.~4,
lower right panel).  Observations of the DFs in this region also show
higher decelerations and lower velocities.  
Fibrils in the dense plage region
are driven by shocks with slightly lower amplitude, since the
photospheric power spectrum peaks at periods of 5 minutes, and in addition 
the strong magnetic fields in dense plage regions generally reduce the
amplitude of convective flows and global oscillations at the
photospheric level. These DFs also experience the component of
gravity along the magnetic field, which in dense plage regions with
more vertical field implies higher decelerations. The 
short lifetimes, high decelerations and low velocities lead to
shorter DFs. 

In region 2, where the field is inclined from the vertical, the
acoustic cutoff period increases to 5 minutes, allowing waves with the
full photospheric peak power at 5 minutes to propagate along the
field into the chromosphere, and develop into shocks that are
stronger than in region 1 and which drive DFs with lifetimes of 5
minutes. Wavelet analysis of region 2 confirms a preponderance of wave
trains with periods of $\sim$ 5 minutes (Fig.~4, lower right panel).
The longer lifetimes and less vertical field lead to the observed
lower decelerations and longer lengths.

\section{Discussion and Conclusion}

There are many striking similarities between quiet Sun mottles and the
active region DFs studied here. Both phenomena appear as highly
dynamic, dark features in the wings and core of H$\alpha$, and are
associated with magnetic flux concentrations. More importantly,
\citet{Suematsu+etal1995} found evidence that quiet Sun mottles also
follow parabolic paths with decelerations that are too small to be
consistent with a purely ballistic flight at solar gravity. While the
interpretation of their observations proved difficult without detailed
numerical models, \citet{Suematsu+etal1995} note that the apparent
velocity profiles in mottles are fully compatible with impulsive
acceleration followed by a constant deceleration, with maximum upward
velocities usually about equal in amplitude to the maximum downward
velocities.  The velocities they report for mottles, of order
10-30~km/s, are similar to those we find in our SST observations of
DFs.  \citet{Suematsu+etal1995} also find that the largest Doppler
velocities in mottles appear at the beginning of the ascending phase
(blue-shifts) and at the end of receding phase (red-shifts), with
downward red-shifted motion sometimes occurring close to their base
during the ascending phase.  These observations agree well with the
properties of our simulated jets (Fig. 2, lower panel). In addition,
\citet{Christopoulou+etal2001} observe limb spicules with clear
parabolic paths with decelerations and maximum velocities similar to
those for mottles and DFs.

All of these strong similarities between previous mottle and spicule
observations and our modeling and SST observations of DFs seem to
imply that highly dynamic chromospheric shock waves cause significant
up- and downward excursions of the upper chromosphere in both active
region and quiet Sun, as proposed by \citet{DePontieu+etal2004}.  Some
unresolved issues remain, such as the longer lifetimes of quiet sun
mottles and spicules (2-10 minutes), and the greater heights
of 2-10 Mm that spicules reach at the limb.  
Preliminary analysis of our simulations suggests that these
differences could be related to large scale differences in magnetic
topology. Further numerical simulations of various magnetic
topologies will help resolve these issues. For example, it is possible
that spicules reach slightly greater heights because they
consist of two populations: jets that are driven by shocks (as
described here), and jets caused by reconnection.  The latter jets
could form a subset that on average is taller than the shock driven
jets, and perhaps be part of a continuous spectrum of reconnection
jets that includes surges, macrospicules and H$\alpha$ upflow events
\citep[]{Chae+etal1998}.  Whatever the role of reconnection in quiet
Sun, our findings indicate that, at least in active regions, most jets
are caused by chromospheric shocks driven by convective
flows and oscillations in the photosphere.

\acknowledgements{This research was supported by NASA grants
  NAG5-11917, NNG04-GC08G and NAS5-38099 (TRACE), the European
  Community's Human Potential Programme contracts HPRN-CT-2002-00313
  and HPRN-CT-2002-00310, The Research Council of Norway through grant
  146467/420 and through grants of computing time from the Programme
  for Supercomputing. The Swedish 1-m Solar Telescope is operated on
  La Palma by the Institute for Solar Physics of the Royal Swedish
  Academy of Sciences in the Spanish Observatorio del Roque de los
  Muchachos of the Instituto de Astrof{\'\i}sica de Canarias. The
  authors are grateful to K.  Schrijver and M. DeRosa for helpful
  discussions. BDP thanks ITA/Oslo for excellent hospitality and GDP
  for well-timed parturition.}

\end{document}